\documentclass[twocolumn,prb,amsmath,amssymb]{revtex4-1}
\usepackage{verbatim}
\usepackage{graphicx}
\usepackage{epstopdf}
\usepackage{amsmath}
\usepackage{amssymb}
\usepackage{wrapfig}
\begin{document}

\title{MOVPE growth, transmission electron microscopy and magneto-optical spectroscopy of individual InAs$_{x}$P$_{1-x}$/Ga$_{0.5}$In$_{0.5}$P quantum dots}

\author{O. Del Pozo-Zamudio$^1$}
\email{o.delpozo@uni-muenster.de. Present address: Institute of Physics, University of M\"{u}nster, 48149 M\"{u}nster, Germany}
\author{J. Puebla$^{1,2}$}
\email{jorgeluis.pueblanunez@riken.jp}
\author{A. Krysa$^3$}
\author{R. Toro$^1$}
\author{A. M. Sanchez$^4$}
\author{R. Beanland$^4$}
\author{A. I. Tartakovskii$^1$}
\author{M. S. Skolnick$^1$}
\author{E. A. Chekhovich$^1$}
\email{e.chekhovich@sheffield.ac.uk}
\affiliation{$^1$Department
of Physics and Astronomy, University of Sheffield, Sheffield S3
7RH, United Kingdom} \affiliation{$^2$Center for Emergent Matter
Science, RIKEN, Wako, Saitama 351 - 0198, Japan}
\affiliation{$^3$Department of Electronic and Electrical
Engineering, University of Sheffield, Sheffield S1 3JD, United
Kingdom} \affiliation{$^4$Department of Physics, University of
Warwick, Coventry, CV4 7AL, United Kingdom}
\date{\today}

\begin{abstract}
We report on growth and characterization of individual
InAs$_{x}$P$_{1-x}$/GaInP quantum dots with variable nominal As
molar fraction. Magneto-photoluminescence experiments reveal QD
emission in a wide range from 1.3 to 1.8~eV confirming
incorporation of As into quantum dots. Transmission electron
microscopy reveals a core-cap structure of InAsP quantum dots with
an InAs-rich core capped by an InP-rich layer. Inside the core, an
As molar fraction up to $x$=0.12 is observed. The heavy hole
$g$-factor is found to be strongly dependent on As molar fraction,
while the electron $g$-factor is close to the InP values. This
suggests type-II carrier confinement in the studied InAsP dots
with holes (electrons) localized in the core (cap) region.
Finally, dynamic nuclear polarization is observed which allows for
further insight into structural properties using nuclear magnetic
resonance.


\end{abstract}

\maketitle

\section{Introduction}
Semiconductor quantum dots (QDs) play a crucial role in emerging
semiconductor device technologies such as single photon sources
and detectors, quantum memories and logic gates
\cite{bib:bimberg}. Their electronic properties can be tailored by
modifying their size and composition. For example, electronic
properties of QDs can be engineered using ternary alloys. In this
context, ternary \textrm{III$_x$III$_{1-x}$V} QD systems have
received most attention. Self-assembled Stranski-Krastanov
In$_x$Ga$_{1-x}$As/GaAs quantum dots are the most studied system,
in which the alloy composition and dot size can be modified to
obtain a broad range of emission energies \cite{bib:MitSug,
bib:Bimberg4.1, bib:Shumway}.

On the other hand, ternary III V$_x$V$_{1-x}$ Stranski-Krastanov
QDs have not been studied in detail. InAs$_{x}$P$_{1-x}$ QDs grown
by self-assembly in Ga$_{0.5}$In$_{0.5}$P is the system considered
in the present work. Due to the significant difference between the
bandgaps of InAs and Ga$_{0.5}$In$_{0.5}$P ($\sim$1.5~eV at room
temperature), a pronounced increase in confinement energy can be
expected for InAsP/GaInP QDs compared to InP/GaInP QDs, favouring
robust performance of QDs at elevated temperatures. The first
report on Stranski-Krastanow growth of InAsP QDs in GaInP by
metalorganic vapour phase epitaxy (MOVPE) was published by
Vinokurov \textit{et al.} \cite{bib:vinokurov}. However, no
significant red-shift of luminescence was observed compared to InP
QDs which could be a result of either inefficient As incorporation
into QDs, or the reduction of QD sizes under As incorporation. By
contrast, Fuchi \textit{et al.} have grown InAsP QDs using droplet
hetero-epitaxy technique and observed a significant red-shift and
broadening of the ensemble QD emission with increased As fraction
\cite{Fuchi20082239}. Ribeiro \textit{et al.} reported experiments
on InAsP/GaAs structures, where quantum dot emission at 77 K was
found to be around 1.25~eV , which lies above the emission energy
of InAs/GaAs measured under the same conditions
\cite{bib:ribeiro}. In the work of Ribeiro \textit{et al.} the
electronic properties of InAsP/GaAs QDs were controlled by the
PH$_3$ flux during the MOVPE process: as the flux was increased,
the QD emission energy increased towards the InP QD energy
\cite{bib:maltez}. In our work we follow a similar approach and
use the flux of AsH$_3$ to control the QD properties. Some recent
examples of QD growth using ternary InAsP alloy also include
demonstration of the InAsP QD lasers~\cite{Karomi,krysa} and
observation of ultraclean emission from InAsP QDs embedded in InP
nanowires~\cite{dalacu}.

To the best of our knowledge, here we present the first report on
growth, transmission electron microscopy and magneto-optical
studies of individual InAsP/GaInP quantum dots, which offer deeper
confinement potential energies compared to the previously studied
InP/GaInP and InAsP/GaAs QDs. Magneto-photoluminescence
(magneto-PL) studies reveal detailed information of the electron
and hole $g$-factor dependence on quantum dot emission energy
which varies in a wide range between 1.3 and 1.8 eV. Such
knowledge of the $g$-factors is key for development of
technologies that employ QD spins \cite{bib:Kosaka,vanbree,witek}.
A combination of results from magneto-PL and transmission
electronic microscopy (TEM) imaging suggests type-II carrier
confinement in the studied InAs$_{x}$P$_{1-x}$ QDs with
sufficiently large As molar fraction $x\sim0.1$. Recently, type-II
QDs have attracted considerable attention as potential candidates
for efficient QD solar cells due to their increased carrier
lifetime and suppressed Auger recombination
\cite{KimS,Laghumavarapu,Ning,Tayagaki}.

The rest of the paper is organized as follows. The details of
sample growth and experimental techniques are described in Sec.
\ref{sec:experimental}. The experimental results are presented and
discussed in Sec. \ref{sec:results}. Finally, in
Sec.~\ref{sec:conclusions} we summarize the results of our work.

\section{Samples and experimental methods} \label{sec:experimental}

Our samples of ternary InAs$_{x}$P$_{1-x}$ QDs embedded in
Ga$_{0.5}$In$_{0.5}$P matrix, were grown by low pressure (150
Torr) MOVPE in a horizontal flow reactor, on (100) GaAs substrates
with a miscut angle of 3$^{\circ}$ towards (1$\bar{1}$0).
Trimethylgallium (TMGa) and Trimethylindium (TMIn) were used as
precursors for group III elements, and arsine (AsH$_3$) and
phosphine (PH$_3$) were used as precursors of group V. Hydrogen
was used  as carrier gas. The GaAs buffer layer and the subsequent
GaInP barrier were grown at 690$^{\circ}$C. The growth rates were
maintained at 0.76 nm/s for the GaAs buffer layers and GaInP
barriers. QDs were deposited at a lower nominal growth rate of
0.11 nm/s. During the growth of the GaInP barriers and quantum
dots the PH$_3$ flow was kept constant at 300~sccm, while the
composition of the InAsP QDs was controlled by the flow rate of
AsH$_3$. Before the deposition of the QDs, the growth was halted,
and the susceptor temperature was lowered to 650$^{\circ}$C. The
growth of the QDs included three steps: The first step involved
deposition of a nominally binary InP for 1~s, this was followed by
InAsP deposition for 3~s during which arsine was introduced to the
reactor, finally a nominally binary InP was grown again for 1~s.
The sample with pure InP/GaInP dots (0~sccm arsine flux) was
produced by growing InP for 5~s in a single step. In what follows
we label the samples by the arsine flux used during the dot
growth.

In order to assess the nominal molar fractions of arsenic in QDs
grown with different AsH$_3$ flows, four InAsP/InP superlattice
(SL) samples have been grown on InP substrates under the same
growth conditions as for QD structures. The growth times of the
InAsP layers were 5~s or 10~s and the total growth times of one
complete SL period were 60~s or 120~s, respectively. The SL
structures were examined by means of X-ray diffractometry as
described in \cite{krysa}. The molar fractions were derived from
the position of the zero order SL peak with respect to the peak
from the InP substrate. The resulting arsenic molar fractions in
the InAsP layers are $x=$0.06, $x=$0.072, $x=$0.084, and $x=$0.108
for AsH$_3$ flows of 5.9~sccm, 10.6~sccm, 16.7~sccm, and 20.2~sccm
respectively.

Optical characterization of QDs was carried out using a
micro-photoluminescence ($\mu$PL) setup equipped with a confocal
low-temperature optical microscope system. An external magnetic
field up to 10 T parallel (Faraday geometry) or perpendicular
(Voigt geometry) to the sample growth axis was applied using a
superconducting magnet. In most experiments PL was excited using
either HeNe laser ($E_\textrm{exc}=1.96$~eV) or diode laser
($E_\textrm{exc}=1.88$~eV), with additional diode lasers
($E_\textrm{exc}=1.80$~eV, $E_\textrm{exc}=1.53$~eV) used for
experiments on nuclear spin effects. Photoluminescence signal was
collected with a 0.85~m double spectrometer and a liquid nitrogen
cooled charge coupled device (CCD). All optical spectroscopy
experiments were carried out at a sample temperature of 4.2~K.
Microscopy characterisation of the QDs presented in section
\ref{ssec:TEM} was conducted using a composition-sensitive
high-resolution transmission electron microscopy (TEM), the
details of this methods are described on Ref. \cite{Karomi}.

\section{Results and discussion} \label{sec:results}

\subsection{Transmission Electron Microscopy} \label{ssec:TEM}

\begin{figure}
\includegraphics[width=0.8\linewidth]{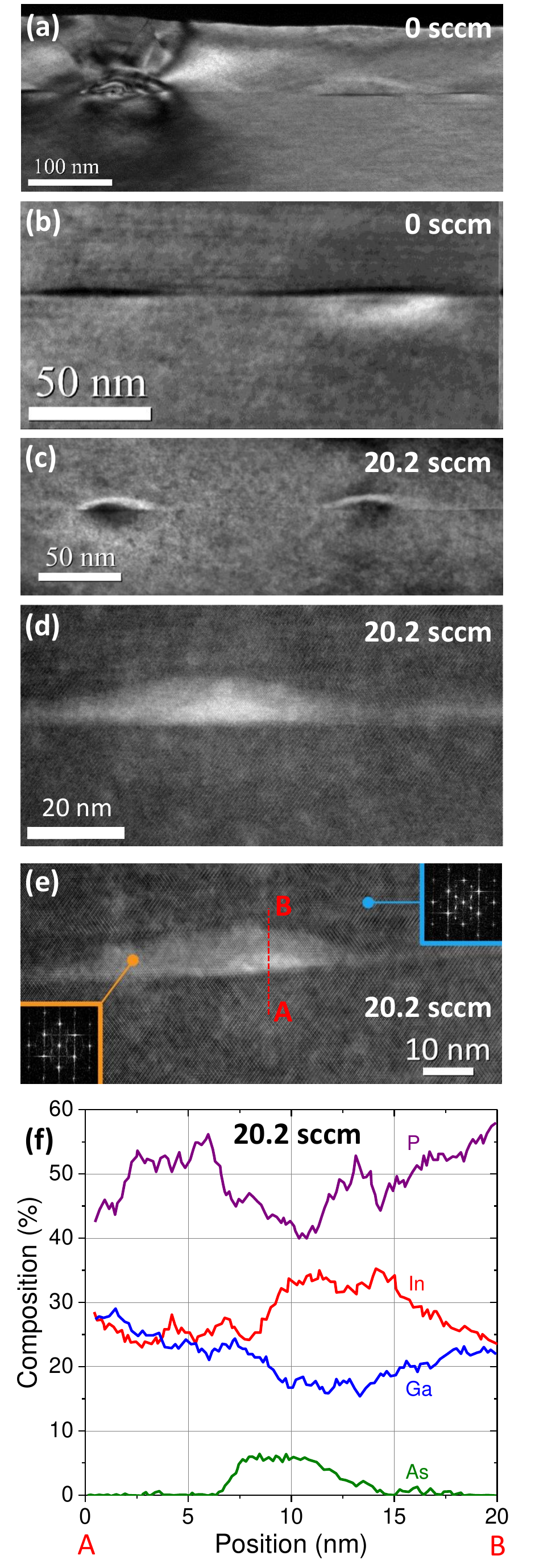}
\caption{Transmission electron microscopy (TEM) of InP/GaInP
quantum dots in 0 sccm sample (a,b) and InAsP/GaInP dots in 20.2
sccm sample (c-f). Images (a)-(c) were taken under 002 dark field
conditions. Images (d,e) were obtained using scanning TEM under
annular dark field (ADF) conditions. Insets in (e) show Fourier
transforms revealing Cu-Pt type ordering in the GaInP barrier, not
observed in a quantum dot. (f) Results of energy-dispersive X-ray
analysis for the quantum dot shown in (e): composition of each
chemical element is shown along the dashed line in (e) in the
direction from point A to point B.} \label{fig:tem}
\end{figure}

We carried out transmission electron microscopy (TEM) studies in
order to examine the morphology and chemical composition of the
quantum dots. Fig.~\ref{fig:tem}(a) shows a conventional TEM image
taken under 002 dark field conditions on a sample grown without As
(0~sccm). Contrast in these images is sensitive to the difference
in the mean atomic number of the group III and V superlattices.
Two types of InP/GaInP quantum dots are observed; large
pyramid-shaped dots (e.g. left side of the image) and smaller disk
shaped dots (e.g. right side of the image). Further examples of
disk shaped InP/GaInP dots are shown in Fig.~\ref{fig:tem} (b).
Similar 002 dark field TEM images of InAsP/GaInP dots (20.2~sccm
sample) are shown in Fig.~\ref{fig:tem}(c). It is apparent that
these InAsP dots are pyramid shaped and have smaller lateral
dimensions of $\sim$40~nm as opposed to $\sim$80 nm for both types
of InP dots in Figs.~\ref{fig:tem}(a, b).

In order to examine the chemical composition, further studies on
InAsP/GaInP dots (20.2~sccm sample) were conducted using
aberration-corrected scanning TEM (AC-STEM). Representative images
taken under annular dark field (ADF) conditions are shown in
Figs.~\ref{fig:tem}(d, e). In these images brighter areas
correspond to elements with larger atomic number $Z$. It is
apparent that a typical InAsP dot consists of a core containing
heavy elements covered by a cap of lighter elements. The image is
aligned with the (001) planes horizontal, showing the QD has
formed preferentially in a local steepening of the 3$^{\circ}$
offcut surface.  Cu-Pt type ordering in the InGaP matrix is
evident from the fast Fourier transform (FFT) shown in top right
box in Fig.~\ref{fig:tem}(e); no such ordering is observed in the
QD core or cap (FFT in bottom left box).

In order to quantify segregation of elements with different $Z$
inside QDs, we have performed energy-dispersive X-ray analysis
(EDX): the results of the scan along the dashed line for the
quantum dot shown in Fig.~\ref{fig:tem}(e) are presented in
Fig.~\ref{fig:tem}(f). It can be seen that the quantum dot
consists of a bottom ''core'' region that is rich in As and In
(which substitute P and Ga of the Ga$_{0.5}$In$_{0.5}$P barrier
respectively), and the top ''cap'' region that is rich in In only.
Taking the arsenic/phosphorus ratio in Fig.~\ref{fig:tem}(f) we
can estimate the arsenic fraction to be $x\approx0.12$, which is
in good agreement with $x=0.108$ derived from X-ray diffractometry
on the reference superlattice sample grown with the same arsine
flow rate of 20.2~sccm.

In summary, TEM imaging shows that the sample growth with AsH$_3$
flow results in arsenic incorporated into quantum dots. The
resulting InAsP dots have notably smaller lateral dimensions than
InP dots and exhibit a core-cap structure, resembling the
core-shell structure of colloidal dots \cite{KimS}. As we show
below, these findings agree with the measurements of diamagnetic
shifts and carrier $g$-factors in individual dots. Furthermore, we
present experimental results that point to type-II confinement in
such core-cap geometry with electrons localized in the InP-rich
cap and and holes occupying the InAs-rich core.

\begin{figure*}
\includegraphics[width=0.9\linewidth]{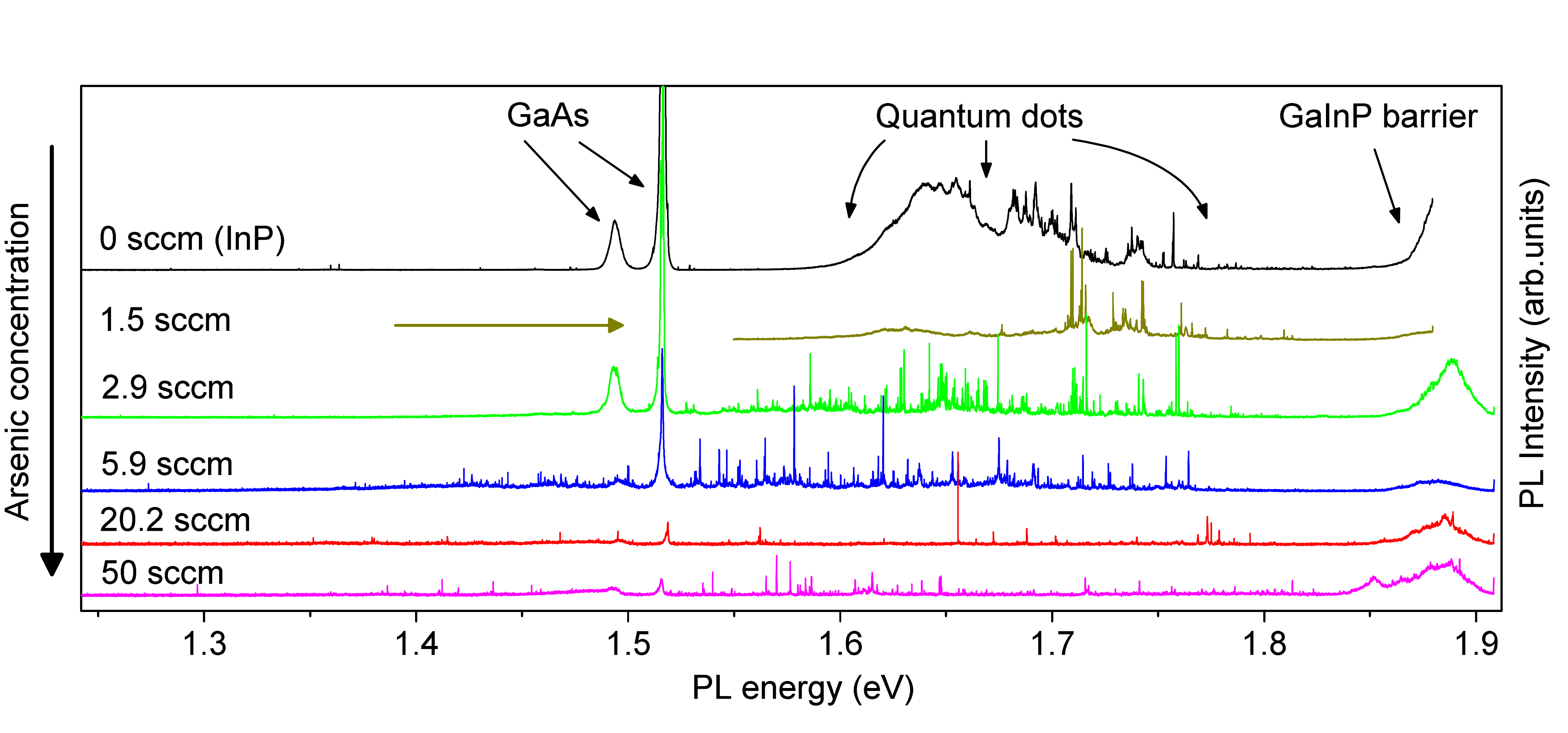}
\caption{Low-temperature photoluminescence spectra of six
InAsP/GaInP QD samples with different nominal As concentration
measured under HeNe laser excitation ($E_\textrm{exc}=1.96$~eV).
The AsH$_3$ flux used for the growth of each sample is given in
standard cubic centimeters per minute (sccm). Emission from GaAs
substrate, quantum dots, and GaInP barrier can be seen in the
spectra. Increased AsH$_3$ flux results in a pronounced red-shift
of the emission, signifying incorporation of As into quantum dots.
This is accompanied by reduction of the quantum dot luminescence
intensity as well as suppression of the GaAs emission which is
attributed to absorption by the quantum dot layer. The variations
in GaInP emission energy and reduction with respect to the values
for disordered bulk material ($\sim$1.99~eV,
Ref.\cite{PhysRevB.47.12598}) are likely due to Cu-Pt ordering, in
agreement with TEM results in shown Fig.~\ref{fig:tem}(e).}
\label{broadspectra}
\end{figure*}

\subsection{Effect of arsenic incorporation on quantum dot photoluminescence} \label{ssec:PL}

Figure \ref{broadspectra} shows the $\mu$PL spectra of six InAsP
QDs samples grown with different nominal As concentrations. The
spectra were measured in a wide range of energies (1.3 - 1.85 eV).
The relative concentrations of As are given in terms of AsH$_3$
flux in standard cubic centimeters per minute (sccm) on the left
side of the graph. The top spectrum (black line) shows QD emission
of the sample without arsenic (InP/GaInP). The spectrum is similar
to those reported previously \cite{Elliott2010}: the sharp peaks
at 1.67 - 1.8 eV are attributed to the disk shaped quantum dots
[see Fig.~\ref{fig:tem}(b)], while broad emission features
centered at 1.65 eV most likely originate from the large pyramidal
quantum dots [see Fig.~\ref{fig:tem}(a)]. When the AsH$_3$ flux is
increased (spectra from top to bottom) the spectral range of
quantum dot emission peaks progressively widens, extending below
the GaAs substrate emission at 1.52 eV for As concentrations above
2.9 sccm. The samples with the largest AsH$_3$ flux (20.2 sccm and
50 sccm) exhibit single-dot emission in a wide range spanning from
1.3 to 1.8 eV. Importantly, there are quantum dots with emission
energies below the bulk band gap of InP (1.421~eV at $T$=4.2~K).
Such pronounced shift of PL emission to lower energies is a clear
sign that arsenic is incorporated into quantum dots.

It follows from the spectra of Fig. \ref{broadspectra}, that
quantum dot PL intensity decreases with increasing arsenic
concentration. As we explain later, we ascribe such behaviour to
the transition from type-I to type-II carrier confinement for
quantum dots with high arsenic concentration.

Even for the highest As fraction, the typical luminescence
linewidths of the studies QDs are less than $\sim$50$\mu$eV
suggesting that As incorporation does not deteriorate the optical
quality of the InAsP/InP dots.

\subsection{Magneto-photoluminescence spectroscopy} \label{ssec:magneto}

In this section we present results of $\mu$PL spectroscopy in
external magnetic fields for quantum dot samples with different As
concentration. Using these data we explore how electron and hole
states are modified by incorporation of As into InP QDs. As we
show, such studies also provide information on the chemical
composition and structure of the InAsP QDs, complementary to TEM
imaging.

\subsubsection{Derivation of the quantum dot charge states and $g$-factors}

\begin{figure*}
\includegraphics[width=\linewidth]{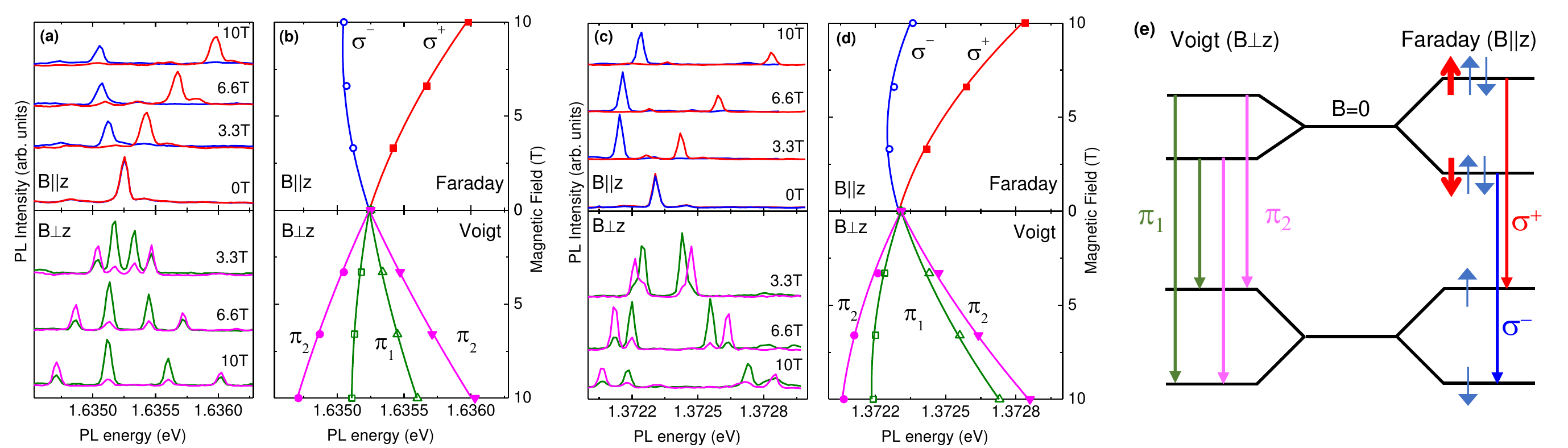}
\caption{(a) Typical magnetic field dependence of PL spectra from
an individual InAsP/GaInP quantum dot in a 20.2 sccm sample
measured at 4.2 K under non-resonant excitation in $\sigma^+$ and
$\sigma^-$ polarized detection in Faraday geometry (top part
$B\parallel z$) and in orthogonal $\pi_1$, $\pi_2$ linearly
polarized detection in Voigt geometry (bottom part $B\perp z$).
The spectral pattern observed in Voigt geometry reveals that
photoluminescence originates from singly-charged dots
\cite{bayer2002}. (b) Energies of the peaks derived from the
spectra in (a) versus external magnetic field (symbols). Solid
lines show the fitting with Equations~\ref{eq:Far} and
\ref{eq:Voigt1} allowing $g$-factors and exciton diamagnetic
shifts to be determined. (c) Magnetic field PL dependence of
another QD from the same sample emitting at lower energy. (d) PL
peak energies from (c) and fitting. (e) Schematic diagram of spin
levels and allowed optical transitions in a negatively charged
InAsP quantum dot in Faraday and Voigt configurations. Electrons
(holes) are shown with thin (thick) arrows representing spin-up
and spin-down states.} \label{fig:specSQD}
\end{figure*}

We first present magneto-PL spectroscopy data which reveals
information on the charge states of the QDs. Two different
geometries were used in our measurements: in Faraday (Voigt)
geometry magnetic field is applied parallel (perpendicular) to the
sample growth axis $B\parallel z$ ($B\perp z$). Figure
\ref{fig:specSQD}(a) shows PL spectra of a single quantum dot in a
sample with large As concentration (20.2~sccm) detected in two
circular polarizations in Faraday geometry and in two orthogonal
linear polarizations in Voigt geometry. Magnetic field dependence
of the spectral positions of the peaks observed in Fig.
\ref{fig:specSQD}(a) is shown in Fig. \ref{fig:specSQD}(b) by the
symbols. Zeeman splitting and diamagnetic shift are observed both
for $B\parallel z$ and $B\perp z$. In Faraday geometry the QD
emission line splits into a circularly polarized ($\sigma^+$ and
$\sigma^-$) doublet, whereas in Voigt geometry a quadruplet of
linearly polarized ($\pi_1$ and $\pi_2$) lines is observed.

In order to describe the dependence of PL peak energies on
magnetic field $B$ we use the following equations
\cite{bayer2002}:
\begin{eqnarray}
E_F(B) &=& E_0 + \kappa_F B^2 \pm \frac{1}{2} g_X \mu_B B \label{eq:Far} \\
E_{V}(B) &=& E_0 + \kappa_V B^2 + \frac{1}{2} \mu_B B (\pm g_{h,\perp}  \pm g_e) \label{eq:Voigt1}
\end{eqnarray}
where $E_0$ is the emission energy at $B=0$, $\mu_B$ the Bohr
magneton, $\kappa_F$ and $\kappa_V$ are the diamagnetic shifts in
Faraday and Voigt geometry, $g_X=(g_{h,\parallel}-g_e$) is the
exciton $g$-factor in Faraday geometry, $g_{h,\parallel}$
($g_{h,\perp}$) is the heavy hole $g$-factor along (perpendicular
to) the sample growth axis, and electron $g$-factor $g_e$ is
assumed to be isotropic. We performed simultaneous least-square
fitting of the data measured in Faraday (Voigt) geometry to
Eq.~\ref{eq:Far} (Eq.~\ref{eq:Voigt1}). The fitting results for
the data in Figs. \ref{fig:specSQD}(a,b) are shown by the lines in
Fig. \ref{fig:specSQD}(b) and yield $g_X$= +1.592,
$g_{h,\parallel}$=+3.175, $g_e$= +1.58, $|g_{h,\perp}|$=0.737,
$\kappa_F$=2.67 $\mu$eV/T$^2$ and $\kappa_V$=1.12 $\mu$eV/T$^2$.
The same analysis is presented in Figs. \ref{fig:specSQD}(c,d) for
another single dot from the same sample emitting at lower energy.
Once again the data is well described by equations
Eqs.~\ref{eq:Far}-\ref{eq:Voigt1} but with notable difference in
$g$-factors and diamagnetic shifts which is discussed in more
detail in Subsections \ref{ssec:diamag} and \ref{ssec:gfactors}.
Due to the $\pm$ signs in Eqs.~\ref{eq:Far}-\ref{eq:Voigt1} there
is potential ambiguity in the signs of the fitted $g$-factors.
However, the signs of $g_{h,\parallel}$, $g_e$ are reliably
established by comparing with the previous studies on neutral
InP/GaInP quantum dots\cite{chekhovich2010} and bulk
InP\cite{Gotschy}. By contrast the sign of $g_{h,\perp}$ is not
defined and only the absolute value $|g_{h,\perp}|$ can be found
from the fitting.

The patterns of spectral components in
Figs.~\ref{fig:specSQD}(a,c) as well as their good description
within the model of Eqs.~\ref{eq:Far}-\ref{eq:Voigt1} prove that
the observed emission arises from singly-charged quantum dots
\cite{bayer2002}: In particular, in Voigt geometry, all four peaks
maintain similar intensities and converge to the same energy in
the limit of $B\rightarrow0$ as opposed to the behaviour of
''dark'' excitons in neutral quantum dots
\cite{bayer2002,bib:Chekhovich3}. The origin of two (four)
spectral peaks in Faraday (Voigt) geometry is illustrated in Fig.
\ref{fig:specSQD}(c) where spin states and optical transitions are
shown schematically for a negatively charged exciton.

All narrow spectral peaks that have been examined, exhibit similar
trion behaviour in all of the studied samples. This suggests that
all of the studied dots emit from a charged state, which can be
ascribed to the combined effect of background doping and optical
excitation above the GaInP barrier bandgap. Distinguishing between
positively and negatively singly charged dots using PL
spectroscopy alone is difficult. However, we note that high
magnetic field ($B=10$~T) applied in Faraday geometry leads to
unequal intensities of the two Zeeman PL components -- a sign of
relaxation between electron or hole spin Zeeman levels. We observe
dots where both high- and low- energy peak becomes dominant in
high field [compare Figs. \ref{fig:specSQD}(a,c)], suggesting that
both positively and negatively charged quantum dots are
encountered in the studied samples.

\subsubsection{Effect of arsenic incorporation on diamagnetic shifts}\label{ssec:diamag}

\begin{figure}
\begin{center}
\includegraphics[width=0.9\linewidth]{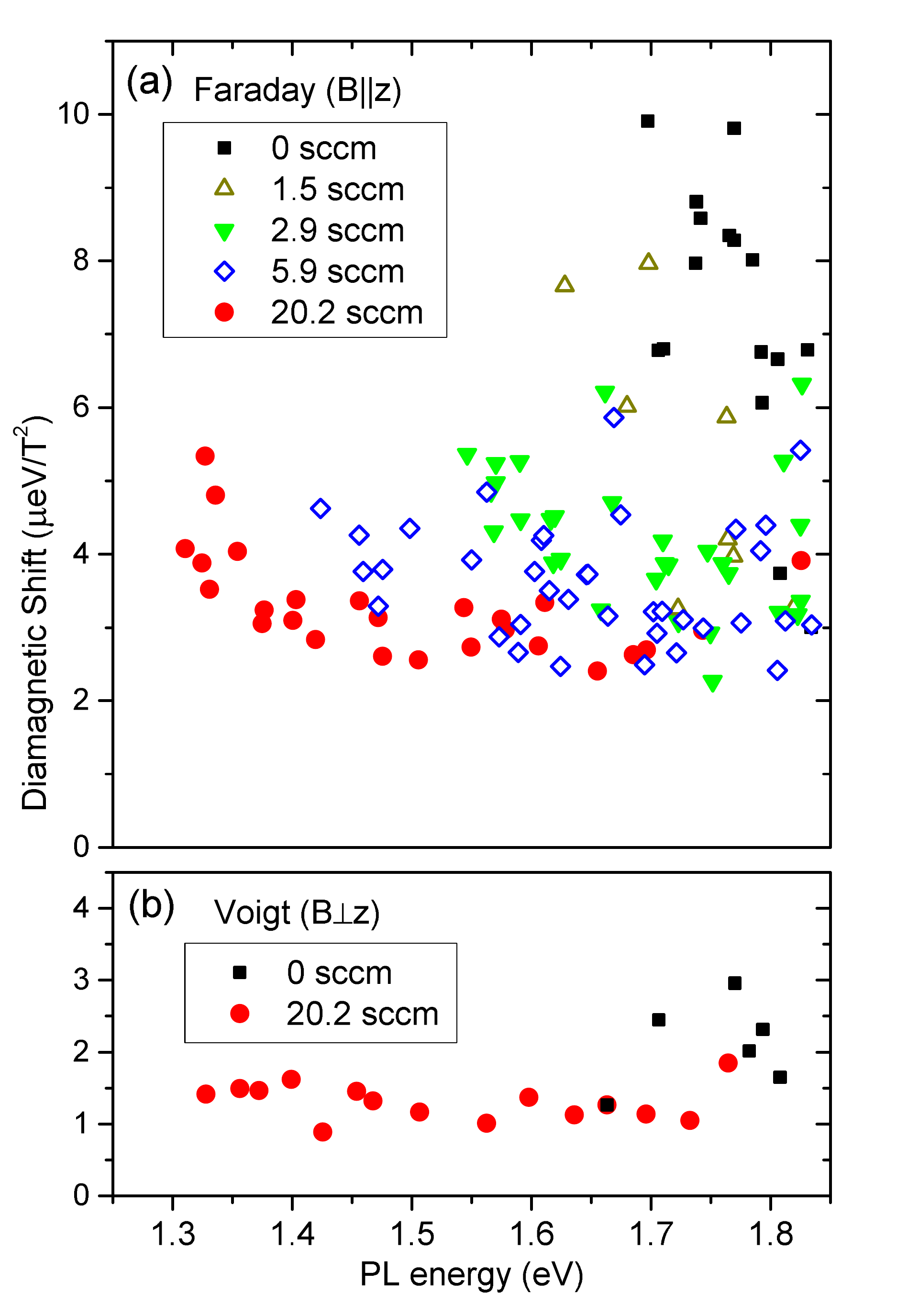}
\caption{Diamagnetic shifts $\kappa$ as a function of emission
energy measured in Faraday (a) and Voigt (b) geometries for a
large number of individual InAsP quantum dots in five samples with
different arsenic concentration (determined by AsH$_3$ flux during
the growth, ranging between 0 and 20.2 sccm). There is a
pronounced reduction of $\kappa_F$ with increasing As fraction
observed in Faraday geometry, revealing the reduction of the
lateral QD sizes induced by As incorporation.}
\label{fig:diamshift}
\end{center}
\end{figure}

Measurements of exciton diamagnetic shifts and exciton $g$-factors
($g_X$) have been carried out for $\sim$120 QDs that have been
selected for relatively bright PL and narrow linewidths in five
samples with different As concentration (AsH$_3$ flux between 0 -
20.2 sccm). Figures \ref{fig:diamshift}(a) and (b) show the
exciton diamagnetic shifts $\kappa_F$ and $\kappa_V$ measured in
Faraday and Voigt geometries respectively. The measurements of
$\kappa_F$ and $\kappa_V$ allow the effect of arsenic
incorporation on quantum dot size and shape to be examined. The
diamagnetic shift $\kappa$ is related to the radius of the exciton
wavefunction $r_X$ in the plane perpendicular to the external
magnetic field by the following equation \cite{bib:walck}:
\begin{equation} \label{eq:diamshift}
\kappa=\frac{e^2}{8 \mu}r^2_X,
\end{equation}
where $e$ is the electron charge and $\mu$ is the reduced exciton
mass.

It can be seen in Fig. \ref{fig:diamshift}(a) that the largest
diamagnetic shifts in Faraday geometry
$\kappa_F\sim$8$\mu$eV/T$^2$ are observed for pure InP dots (0
sccm sample). Increased arsenic concentration results in reduced
$\kappa_F$ for the dots emitting at the same energies. This trend
in diamagnetic shifts suggests that incorporation of arsenic into
InAsP quantum dots results in reduction of their lateral
dimensions. Such conclusion agrees with the TEM results presented
in Section \ref{ssec:TEM}. Furthermore in the sample with large As
fraction (20.2 sccm) $\kappa_F$ tends to increase for quantum dots
with smaller emission energy suggesting their increased lateral
dimensions. The diamagnetic shifts in Voigt geometry $\kappa_V$
presented in Fig.~\ref{fig:diamshift}(b) are notably smaller than
$\kappa_F$ agreeing with the disk-shape nature of the dots
revealed by TEM.

\subsubsection{Effect of arsenic incorporation on $g$-factors and analysis of carrier confinement}\label{ssec:gfactors}

In order to gain deeper insight into the spin properties of InAsP
QDs, we extract the magnitudes of $g$-factors for samples with
different arsenic content. As we show below, these data provides
useful information about quantum dot composition and structure and
suggests type-II carrier confinement.

\begin{figure}
\begin{center}
\includegraphics[width=\linewidth]{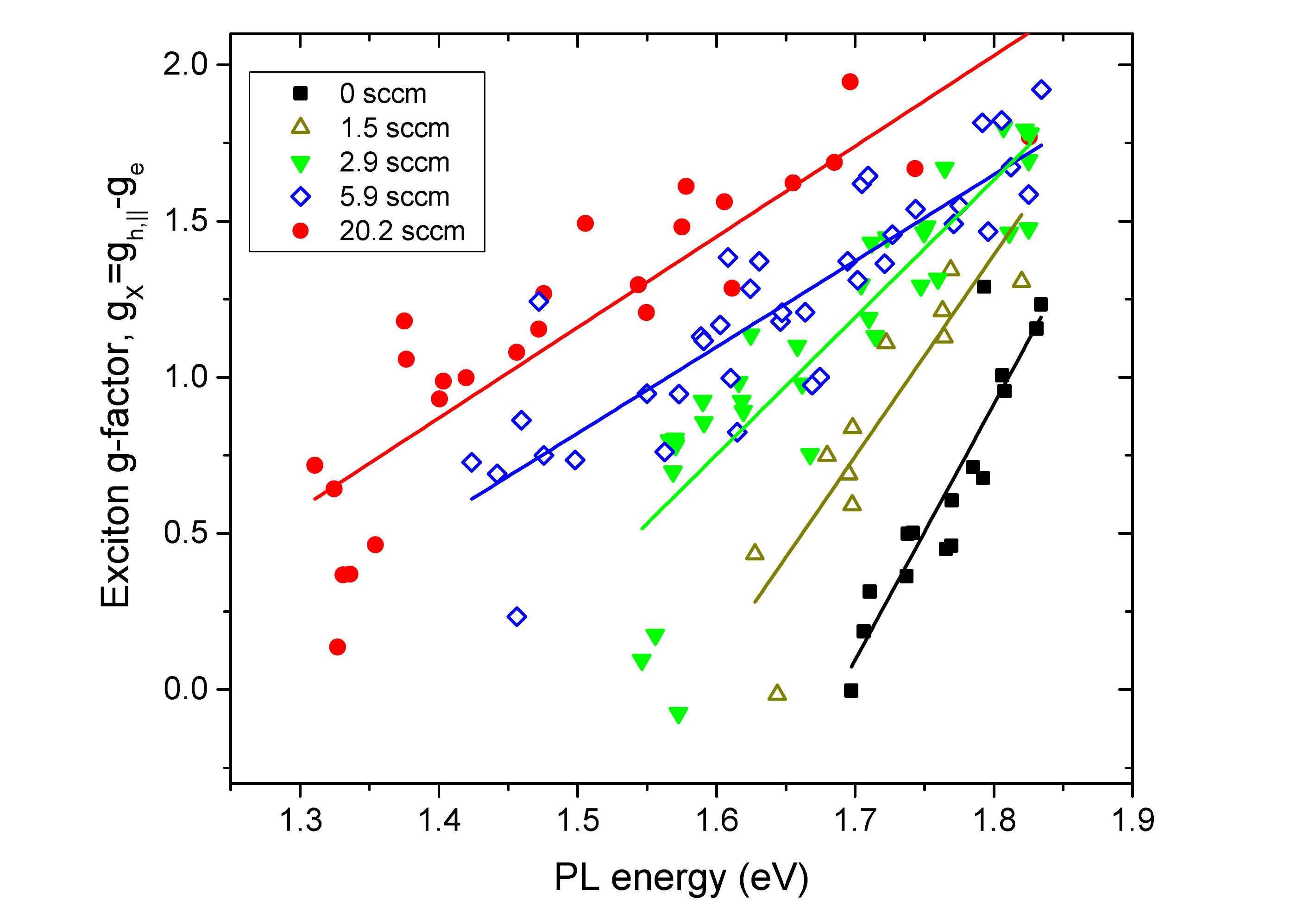}
\caption{Exciton $g$-factors $g_X$ measured in Faraday
configuration as a function of emission energy for a set of
quantum dots five samples with different As concentration. Solid
lines are linear fits (see fitted parameters in Table I).}
\label{fig:gX}
\end{center}
\end{figure}

The symbols in Figure~\ref{fig:gX} show the out-of-plane exciton
$g$-factors $g_X$ measured in Faraday geometry as a function of
the QD ground state emission energy in five samples. Linear fits
are plotted by the lines to visualize the trends in exciton
$g$-factors for samples with different As compositions; the
fitting parameters are listed in Table~I. It can be seen from
Fig.~\ref{fig:gX} that increased AsH$_3$ flux leads to systematic
increase in exciton $g$-factors at all energies $E$ where QD
luminescence is observed. Furthermore, the slopes $m$ of the
$g_X(E)$ dependencies decrease for large As concentration. These
pronounced variations in $g_X$ signify the change in the chemical
composition in the QD volume where exciton wavefunction is
localized and thus confirm successful incorporation of As into the
dots.

\begin{table} \label{tab:lingfit}
\begin{tabular}{l | c | c | c}
Sample & m (eV$^{-1}$) & $g_X$ at $E=1.8$~eV\\
\hline
0 sccm & 8.18567 & 0.904\\
1.5 sccm & 6.459 & 1.381\\
2.9 sccm & 4.400 & 1.640\\
5.9 sccm & 2.758 & 1.640\\
20.2 sccm & 2.899 & 2.012
\end{tabular}
\caption{Parameters derived from the fitting exciton $g$-factors
using linear function $g_X(E)=m (E-1.8~\textrm{eV})+
g_X(1.8~\textrm{eV})$, where $E$ is the QD ground state emission
energy (the fits are shown by the lines in Figure \ref{fig:gX}).}
\end{table}

\begin{figure}
\begin{center}
\includegraphics[width=\linewidth]{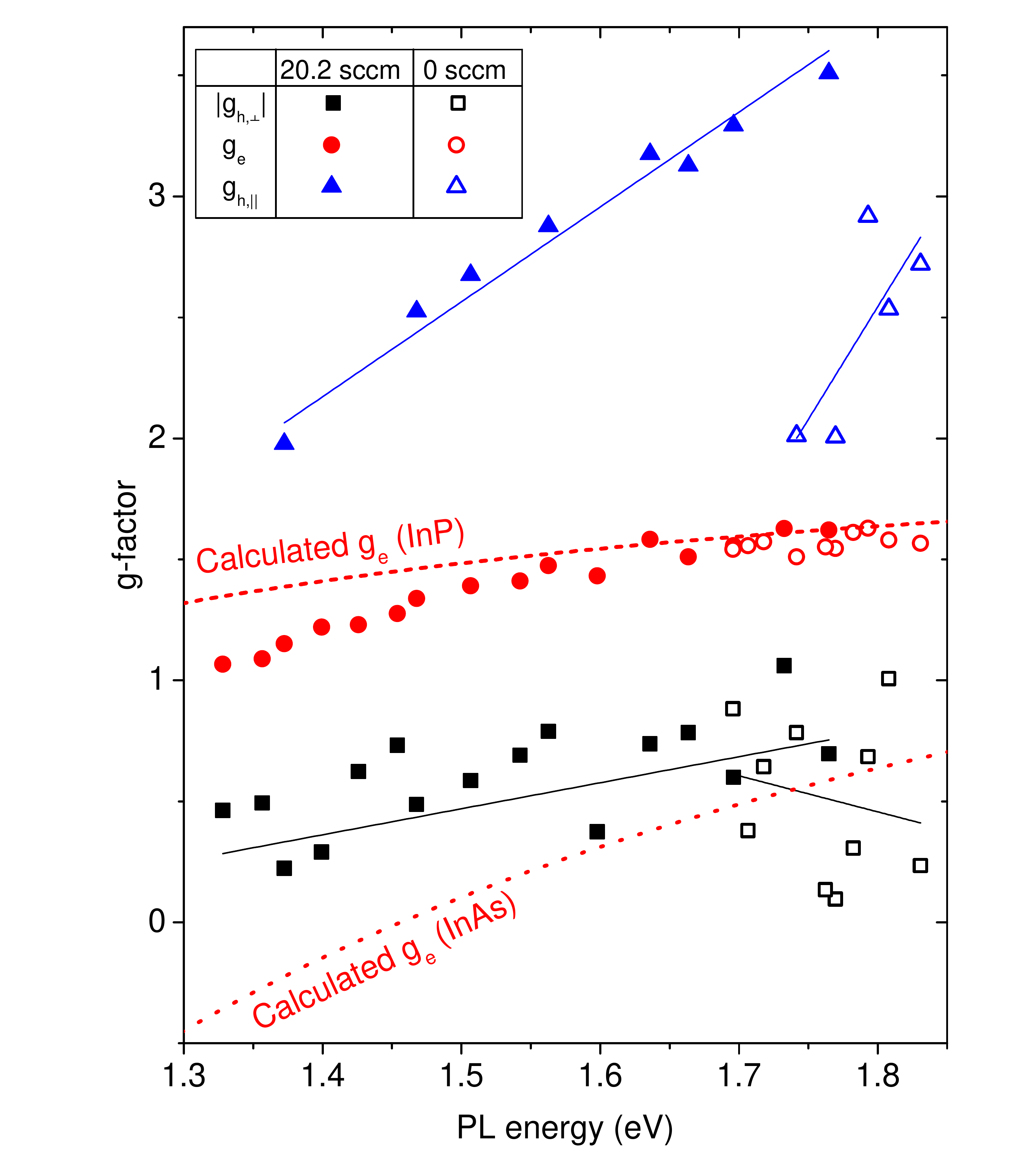}
\caption{Electron g-factors ($g_e$, circles), out-of-plane heavy
hole g-factors ($g_{h,\parallel}$, triangles), and in-plane heavy
hole g-factors ($g_{h,\perp}$, squares) measured for InP/GaInP
quantum dots (0 sccm sample, open symbols) and InAsP/GaInP dots
(20.2 sccm sample, solid symbols) shown as a function of
photoluminescence energy. Thin solid lines show linear fits that
can be used as a guide to an eye. Electron g-factors calculated
using Eq.~\ref{eq:roth} are shown by the dashed line for the case
of InP and by the dotted line for the case of InAs. Very good
agreement between experimental $g_e$ values and theory for InP is
found, suggesting that electron and hole wavefunctions are
spatially separated with electrons localized in the InP-rich areas
of the dots.} \label{fig:gegh}
\end{center}
\end{figure}

In order to gain further insight we examine the contributions of
the electron and hole $g$-factors to the variation of the exciton
$g_X$ observed in Figure \ref{fig:gX}. For this purpose we focus
on two samples with zero and large (20.2 sccm) arsenic
concentration and conduct magneto-PL measurements where the same
dots are measured both in Faraday and Voigt geometry allowing
electron and hole $g$-factors to be derived as described in
Subsection~\ref{ssec:PL}. Figure \ref{fig:gegh} shows the
extracted $g$-factors for the 0 sccm sample (open symbols) and the
20.2 sccm sample (solid symbols). The thin solid lines are linear
fits which can be used as guides to an eye. We first note the
large spread in the $g_{h, \perp}$ values, which is expected since
heavy hole in-plane $g$-factors depend strongly on the anisotropy
of shape and strain of each particular dot \cite{bayer2002}. The
results for out-of-plane heavy hole $g$-factors $g_{h, \parallel}$
and for electron $g$-factors $g_{e}$ are more robust and show a
striking difference: while $g_{e}$ values follow the same trend
for both samples, there is a pronounced deviation in $g_{h,
\parallel}$ values. It is thus evident that it is the heavy-holes
which are the most sensitive to incorporation of arsenic into the
InAsP dots, whereas electrons have similar properties in
structures with and without arsenic.

For quantitative analysis we use the result of Roth et al
\cite{bib:roth} that electron $g$-factor is determined mainly by
the bandgap of the semiconductor. While it was derived originally
for bulk materials this result has been extended successfully to
quantum wells and quantum dots \cite{bib:sirenko, bib:yugova,
bib:syperek}. Thus we write for electron $g$-factor:
\begin{equation} \label{eq:roth}
g_e=2-\frac{2 E_P \Delta}{3 E_g (E_g+\Delta)},
\end{equation}
where $E_g$ is the band gap, $\Delta$ is the spin-orbit splitting
and $E_P$ is the Kane energy parameter. (We use $\Delta$=0.38~eV,
$E_P$=21.11~eV for InAs and $\Delta$=0.11~eV, $E_P$=17.0~eV for
InP as reported in the literature \cite{bib:bastard,
bib:cardona}.) Equation~\ref{eq:roth} is plotted in
Fig.~\ref{fig:gegh} by the dashed line for InP and by the dotted
line for InAs. The simple theoretical equation is in excellent
agreement with experimental $g_e$ values if we assume pure InP
parameters. We thus conclude that both in InAsP and InP dots the
electron behaves as if the dot consists of nearly pure InP, with
small deviation developing only for the dots with the lowest
ground state energy. While $g_{h,
\parallel}$ can not be calculated in a simple way, it is generally
proportional to the $\kappa$ and $q$ parameters of the valence
band \cite{PhysRevB.4.3460} and can thus be expected to be larger
for InAs than for InP. Therefore, the increase in $g_{h,
\parallel}$ observed for the 20.2 sccm sample in
Fig.~\ref{fig:gegh} is attributed to the increased fraction of
arsenic ''sampled'' by the hole wavefunction in the InAsP dots. To
summarize, our observations strongly suggest that electrons and
holes are localized in spatially separated parts of the quantum
dot.

We propose the following interpretation that agrees with the
observed $g$-factor values, reduced luminescence intensity of QDs
with large As concentration, and TEM results of Section
\ref{ssec:TEM}: The growth conditions favour the formation of
core-cap QDs where InP-rich cap region localizes the electron and
is separated from the InAs-rich core region where the hole is
predominantly localized. Such spatial separation of electrons and
holes may give rise to type-II QD behaviour. This conclusion is
non-trivial since electronic band alignment at InAsP/InP interface
is expected to be of type-I (Ref. \cite{JAP35Semicon}). The most
likely explanation is that large inhomogeneous strain
characteristic of self-assembled QDs can lead to significant shift
in energy levels \cite{Grundmann1995} and can be responsible for
type-II band alignment.

Recently, type-II QDs have received increased attention as
promising candidates for QD solar cells applications, where
spatial separation of electrons and holes reduces spontaneous
recombination and favours carrier extraction. Moreover, the
structures studied here exhibit InAsP QDs with a broad range of
the band-gap energies (from $\sim$1.3 eV to $\sim$1.8 eV) which
could be advantageous for light conversion efficiency.

\subsubsection{Optical control of the quantum dot nuclear spins} \label{ssec:nuclear}

All isotopes of the elements present in InAsP/GaInP quantum dots
have non-zero nuclear spins, as a result electron-nuclear
interactions are significant
\cite{NatMatReview,Urbaszek,bib:Abragam}. Using circularly
polarized optical excitation it is possible to inject
spin-polarized electrons into quantum dot. Spin polarized electron
can then transfer their polarization to one of the nuclear spins
of the dot via the electron-nuclear hyperfine interaction.
Repeated optical recombination and re-excitation of the spin
polarized electrons can then lead to substantial polarization of
the quantum dot nuclear spin ensemble, typically consisting of
$\sim$10$^5$ nuclei. Such dynamic nuclear polarization (DNP)
process has been reported previously for different types of
quantum dots \cite{GammonPRL2001,bib:Eble, puebla2013} including
InP/GaInP quantum dots
\cite{bib:Skiba,chekhovich2010,bib:Chekhovich1,bib:Chekhovich2,bib:Chekhovich3}.
Here we extend these studies to InAsP/GaInP quantum dots.

The measurements were conducted on the 20.2 sccm sample in
external magnetic field parallel to the sample growth direction
(Faraday geometry). Magnetic field splits the QD emission peak
into a Zeeman doublet [see Figs.~\ref{fig:specSQD}(a,c)]. Since
the two peaks of the spectral doublet originate from electron
states with opposite spins [see Fig.~\ref{fig:specSQD}(d)] the
onset of nuclear spin polarization results in hyperfine
(Overhauser) shift, i.e. increase or decrease of the Zeeman
splitting $\Delta E$ depending on the direction of the effective
nuclear field. (Here for simplicity we neglect the interaction of
the hole spin with the nuclei since its contribution is at least
10 times smaller than that of the electron\cite{bib:Fallahi,
bib:Chekhovich2}.)

\begin{figure}
\begin{center}
\includegraphics[width=\linewidth]{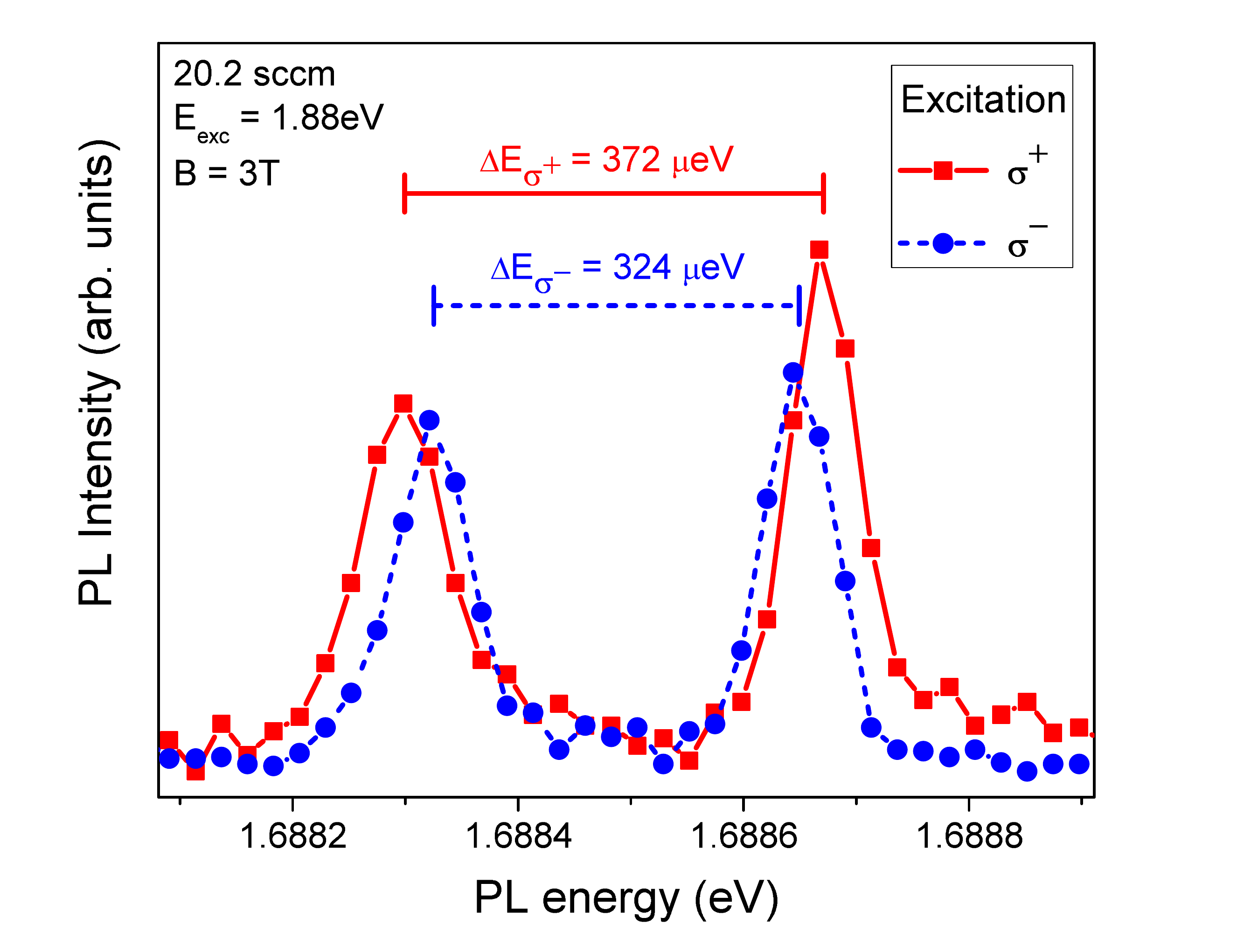}
\caption{Photoluminescence spectra of an individual InAsP/GaInP
quantum dot (20.2 sccm sample) measured under $\sigma^+$ (red
solid line and squares) and $\sigma^-$ (blue dashed line and
circles) circularly polarized optical excitation at $B = 3.0$~T in
Faraday geometry. Variation of the Zeeman doublet splitting
$\Delta E_{\sigma^{\pm}}$ in the trion spectra measured under
$\sigma^{\pm}$ excitation reveals dynamic nuclear spin
polarization. The Overhauser shift for this measurement is
estimated to be $E_\textrm{OHS} = (\Delta E_{\sigma^+} - \Delta
E_{\sigma^-})/2\approx 24~\mu$eV. The Zeeman splittings $\Delta
E_{\sigma^+}$ and $\Delta E_{\sigma^-}$ are shown by the
horizontal bars.} \label{fig:OHS}
\end{center}
\end{figure}

The change in $\Delta E$ induced by DNP can be detected in the PL
spectra as demonstrated in Fig.~\ref{fig:OHS} where two spectra of
the same InAsP/GaInP quantum dot are shown for $\sigma^+$
(squares) and $\sigma^-$ (circles) excitation at $B=3$~T. Gaussian
fitting is used to determine the corresponding Zeeman splittings
$\Delta E_{\sigma,+}$, $\Delta E_{\sigma,-}$ (shown by the
horizontal bars). The Overhauser energy shift $E_\textrm{OHS}$ can
be quantified by the difference between the observed Zeeman
splitting $\Delta E$ and $\Delta E$ corresponding to zero nuclear
polarization ($E_\textrm{OHS}=0$). Realizing the
$E_\textrm{OHS}=0$ condition is demanding, so it is more practical
to estimate the Overhauser shift as $E_\textrm{OHS}\approx(\Delta
E_{\sigma^+}-\Delta E_{\sigma^-})/2$. This equation gives a lower
bound estimate which is exact in case $\sigma^{\pm}$ excitation
produces $E_\textrm{OHS}$ of the same magnitude but opposite
signs. For the measurement presented in Fig.~\ref{fig:OHS} we find
$E_\textrm{OHS}\approx24\mu$eV.

\begin{figure}
\begin{center}
\includegraphics[width=\linewidth]{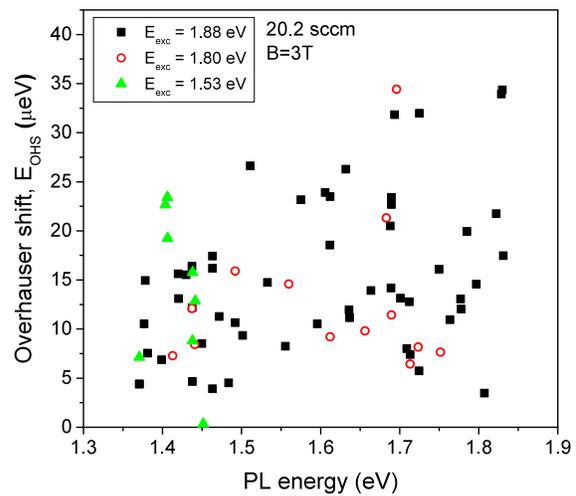}
\caption{Overhauser shifts measured in InAsP/GaInP quantum dots
(20.2 sccm sample) at $B$ = 3 T using circularly polarized
excitation at three different photon energies:
$E_\textrm{exc}$=1.88~eV (squares), $E_\textrm{exc}$=1.80~eV
(circles) and $E_\textrm{exc}$=1.53~eV (triangles).}
\label{fig:OHS2}
\end{center}
\end{figure}

Measurements of the Overhauser shifts were repeated at $B=3$~T on
a set of individual InAsP quantum dots emitting in a wide range of
energies 1.3 - 1.8~eV as shown in Fig.~\ref{fig:OHS2}. Circularly
polarized excitation at different photon energy $E_\textrm{exc}$
was employed. Excitation at $E_\textrm{exc}$=1.88~eV (squares) is
in resonance with the low-energy tail of the GaInP barrier, and
was previously used to induced DNP in InP/GaInP QDs
\cite{bib:Skiba, bib:Chekhovich3}. In addition we used
$E_\textrm{exc}$=1.80~eV (circles) and $E_\textrm{exc}$=1.53~eV
(triangles) to study DNP in QDs emitting at lower energies. DNP
with comparable Overhauser shifts $E_\textrm{OHS}$ is observed for
all $E_\textrm{exc}$ used here. Overall, larger $E_\textrm{OHS}$
is observed for InAsP dots with larger emission energy that are
more reminiscent of InP dots. Nevertheless, the largest
$E_\textrm{OHS}\approx35~\mu$eV observed here for InAsP dots is
significantly smaller than in InP dots, where $E_\textrm{OHS}$
exceeding 120~$\mu$eV has been achieved \cite{bib:Chekhovich1}.
Since P and As have similar nuclear magnetic moments, such
reduction of $E_\textrm{OHS}$ implies smaller degree of the
optically induced nuclear spin polarization in InAsP dots. On the
other hand, such reduction in DNP efficiency agrees with our
hypothesis about the type-II nature of the studied dots: longer
exciton lifetimes can create a bottleneck and lower the efficiency
of the cyclic nuclear spin polarization process
\cite{bib:Chekhovich1,NatMatReview,Urbaszek}. Despite the smaller
$E_\textrm{OHS}$ values, observation of pronounced DNP opens the
way for future studies using optically detected nuclear magnetic
resonance (NMR) spectroscopy \cite{bib:Chekhovich5,
chekhovich2015,chekhovich2013NatPhys,MunschNMR,waeber2016}, which
can provide further insights into chemical composition and strain
profiles in the studied InAsP/GaInP quantum dots.

\section{Conclusions} \label{sec:conclusions}

We have presented a detailed study of individual InAsP/GaInP
quantum dots in samples with different arsenic content grown by
MOVPE. Our samples show QD emission in a broad optical spectral
range (1.3 eV to 1.8 eV), confirming successful incorporation of
arsenic into the dots. The combined analysis of electron
microscopy imaging, exciton diamagnetic shifts and carrier
$g$-factors measurements suggests that InAsP/GaInP dots exhibit
type-II carrier confinement, where holes are localized in the
InAs-rich core, while electrons reside in the InP-rich cap region.
These properties make InAsP quantum dots of interest for efficient
solar cell applications. Future work will include direct
investigation of the type-II confinement by probing electron-hole
recombination dynamics as well as further structural studies
assisted by optically detected nuclear magnetic resonance
techniques.

\section*{Acknowledgments}

This work has been supported by the EPSRC Programme Grant
EP/J007544/1. O. Del Pozo-Zamudio and J. Puebla gratefully
acknowledge the support from CONACYT-Mexico Doctoral Scholarship
programme. E. A. Chekhovich was supported by a University of
Sheffield Vice-Chancellor's Fellowship and a Royal Society
University Research Fellowship.


%

\end{document}